\documentclass[aps, showpacs,twocolumn]{revtex4}

\usepackage{amssymb}
\usepackage{amsmath}
\usepackage{graphicx}
\input{epsf}
\usepackage{epsf}

\begin{document}

\title{Holographic Dark Energy with Curvature}

\author{V\'ictor H. C\'ardenas}

\email{victor[at]dfa.uv.cl}

\author{Roberto G. Perez}

\email{genaro.perez[at]gmail.com}

\affiliation{Departamento de F\'{\i}sica y Astronom\'ia, Universidad
de Valpara\'iso, Gran Breta\~na 1111, Valpara\'iso, Chile}

\begin{abstract}
In this paper we consider an holographic model of dark energy, where
the length scale is the Hubble radius, in a non flat geometry. The
model contains the possibility to alleviate the cosmic coincidence
problem, and also incorporate a mechanism to obtain the transition
from decelerated to an accelerated expansion regime. We derive an
analytic form for the Hubble parameter in a non flat universe, and
using it, we perform a Bayesian analysis of this model using SNIa,
BAO and CMB data. We find from this analysis that the data favored a
small value for $\Omega_k$, however high enough to still produce
cosmological consequences.
\end{abstract}


\maketitle

\section{Introduction}

One of the most important problems in cosmology is to explain the
fact that the universe today is in a phase of accelerated expansion.
In 1998, two teams studying distant Type Ia supernovae presented
independent evidence that the expansion of the Universe is speeding
up \cite{Riess98,Perlmuter99}. The physical origin of cosmic
acceleration remains a deep mystery. Since the expansion is speeding
up, we are faced with two possibilities, either of which would have
profound implications for our understanding of the cosmos and the
laws of physics. The first is that 75 \% of the energy density of
the Universe exists in a new form with large negative pressure,
called dark energy. The other possibility is that General Relativity
breaks down on cosmological scales and must be replaced with a more
complete theory of gravity. Among the models proposed so far for
dark energy, the Holographic Dark Energy (HDE) scenario emerge as
one of the few that has a deep physical substratum.

The idea of Holography was first suggested by Bekenstein
\cite{bekenstein}, by generalizing the results obtained from black
hole physics, and postulating that the maximum entropy in a box of
volume $L^3$ should growth as the \textit{area} of the box. In this
context, 'tHooft \cite{thooft} and Susskind \cite{susskind} stressed
that this principle has to be considered as a recipe for a quantum
gravity theory. The idea soon finds its way in cosmology
\cite{HPple}. The Holographic Principle can be described as follows:
if we want to reconcile quantum mechanics with gravity, we have to
assume that the observable degrees of freedom of the universe are
projections coming from a two-dimensional surface, where the
information is stored.

Considering that particle physics can be successfully described by
an effective field theory with an ultraviolet (UV) cutoff, and black
hole physics impose an infrared (IR) cutoff to the energy density
involved, Cohen et al.\cite{cohen} suggested that the dark energy
should satisfy this principle, proposing a relationship between
cutoff as
\begin{equation}\label{eq1}
L^3 \rho_{\Lambda} \leq L M_p^2,
\end{equation}
where $\rho_{\Lambda}$ is the dark energy density, $L$ the size of
the system, and $M_p=(8\pi G)^{-1/2}$ is the reduced Planck mass. In
\cite{thomas} the author notice that using $L$ as the size of the
observed universe ($L_{today} \simeq 10$ Gy) yields a value
$\rho_{\Lambda} \simeq 10^{-10}$ eV$^4$, which would explain the
observed value of the dark energy density. However, Hsu \cite{hsu}
demonstrated that although identifying $L$ with the Hubble radius
$H^{-1}$ leads to a correct order of magnitude for the dark energy
density, written $\rho_{\Lambda}=3c^2H^2M_p^2$ saturating
(\ref{eq1}) with $c$ constant, this leads to a wrong equation of
state. Based on this, Li \cite{li} proposed instead to identify $L$
with the future event horizon.

Soon after, Pavon and Zimdahl \cite{pav-zim05} showed that is still
possible to consider the Hubble radius as a length scale in the
context of interacting dark energy models \cite{interaction}, which
allows, as a bonus, to alleviate the \textit{cosmic coincidence
problem}: why the energy density of matter and dark energy are of
the same order of magnitude just today. This time however,
$\rho_{\Lambda}=3c^2H^2M_p^2$ can actually describe the dark energy
density assuming that $c$ is a time dependent function. The analysis
performed in that paper was extended in \cite{zim-pav07} for
non-flat cosmologies, focusing in the evolution of the ratio of the
energy densities $r \equiv \rho_m/\rho_{\Lambda}$. An statistical
analysis of this model in a flat universe was performed recently in
\cite{xu09}.

In this paper we consider an interacting holographic dark energy
model, using the Hubble radius as the length scale, in a universe
with arbitrary curvature. The model not only contains the
possibility to alleviate the coincidence problem, but also
incorporate a mechanism to obtain the transition from decelerated to
an accelerated expansion regime. We are primary interested in study
the effects of the inclusion of a small curvature in the theoretical
parameters in light of the observational data. We derive an analytic
form for the Hubble parameter in a non flat universe, and using it,
we perform a Bayesian analysis of this model using supernovae data,
Baryon Acoustic Oscillation (BAO) distance, and the CMB shift
parameter. We find from this analysis that the data favored a small
value for $\Omega_k$, however high enough to produce important
cosmological consequences \cite{GEllis}.

In the next section we review the model clarifying the assumptions
made in \cite{zim-pav07} and \cite{xu09}. In section III we derive
the flat universe results, some of them well known, and others new.
The main contribution is explained in section IV where is derived
the general solution for the non flat case. Next, we perform the
bayesian analysis using observational data, and we end the paper
with some comments.

\section{An effective equation of state for dark energy}

The Einstein field equations for a homogeneous and isotropic
universe are the Friedman equation
\begin{equation}\label{friedman}
H^2+{k \over a^2}=\frac{1}{3M_p^2} \rho,
\end{equation}
where $\rho$ is the energy density, $H=\dot{a}/a$ and $a(t)$ is the scale
factor, $k=+1,0,-1$ for a closed, flat, and open geometries respectively,
and
\begin{equation}\label{hdot}
\dot{H}=-{1 \over 2 M_p^2}(\rho + P) + {k \over a^2},
\end{equation}
where $P$ is the pressure, and where a dot means cosmic time
derivative. The model emerges by assuming that the energy density
$\rho$, which satisfy the usual energy conservation equation
\begin{eqnarray}\label{energycons}
\dot{\rho}+ 3H(\rho + P)=0,
\end{eqnarray}
can be separated into two components; dark matter and dark energy,
so ${\rho}={\rho}_{m}+{\rho}_{\Lambda}$ and $P = p_m+{p}_{\Lambda}$.
Assuming explicitly the equation of state for dark matter,
$p_{m}=0$, and for dark energy $p_{\Lambda}=
w_{\Lambda}{\rho}_{\Lambda}$, and further, considering that neither
component conserves separately, we obtain
\begin{eqnarray}
\dot{\rho}_{\Lambda}+3H\rho_{\Lambda}(1+w_{\Lambda}) = -Q,\label{rol} \\
\dot{\rho}_{m}+3H\rho_{m}=+Q, \label{rom}
\end{eqnarray}
where $Q$ is an arbitrary interaction term between dark matter and dark
energy. From (\ref{rol}) we can take the $Q$ term to the left hand side
and write
\begin{eqnarray}
\dot{\rho}_{\Lambda}+3H(\rho_{\Lambda}+p_{\Lambda}+\frac{Q}{3H})=0,
\end{eqnarray}
which looks like the energy conservation equation for dark energy but
with an effective pressure given by
\begin{eqnarray}
p^{eff}_{\Lambda}=p_{\Lambda}+\frac{Q}{3H}.
\end{eqnarray}
Then, we can write an effective equation of state for this dark energy
as
\begin{eqnarray}\label{weff}
\omega^{eff}_{\Lambda}=\frac{p^{eff}_{\Lambda}}{\rho_{\Lambda}} =
w_{\Lambda}+\frac{Q}{3H\rho_{\Lambda}}.
\end{eqnarray}
We can also rewrite this equation by using (\ref{rol}) and (\ref{rom})
obtaining
\begin{eqnarray}\label{weff1}
\omega^{eff}_{\Lambda}=-1-\frac{1}{3}\frac{d\ln\rho_{\Lambda}}{d\ln
a},
\end{eqnarray}
independent of the value of $w_{\Lambda}$.

\section{The Holographic Model for a flat Universe}

As a warm up and also to introduce the formalism, let us discuss the
flat model first. This section is a critical review of previous
works \cite{pav-zim05}, \cite{zim-pav07} y \cite{xu09}.
\subsection{Constant c}

As we stated in the introduction, the corresponding vacuum
energy density can be written from the holographic dark energy as
\begin{eqnarray}
\rho_{\Lambda}=3c^{2}M^{2}_{p}H^{2},
\end{eqnarray}
where $c$ is considered constant. Combining this with the Friedman
equation enable us to write the energy density of dark matter
\begin{eqnarray}
\rho_{m}=3M^{2}_{p}H^{2}(1-c^{2}).
\end{eqnarray}
To protect the positivity of the dark matter energy density, $c^2<1$
is required. Clearly, if $c$ is constant and the two components
evolve independently (i.e. interaction $Q=0$), then $\rho_m \simeq
H^2 \simeq a^{-3}$ as the usual dark matter, but this implies also
that $\rho_{\Lambda} \simeq a^{-3}$, an evolution that can not
generate accelerated expansion. This is the result found by Hsu
\cite{hsu}, \cite{li}.

However, as first stated in \cite{pav-zim05}, if we consider an
interaction between dark matter and dark energy (see Eqs.(\ref{rol})
and (\ref{rom})), we can obtain an equation of state for dark energy
different from that of dust. This is exactly what we have derived in
the previous section in Eq.(\ref{weff1}). Because $H=\dot{a}/a$ and
using Eq.(\ref{energycons}), we can find the dark energy density as
a function of the scale factor
\begin{equation}\label{rdef}
\rho_{\Lambda}=\frac{c^2}{1-c^2}\rho_{m}\sim a^{-3(1-c^2)},
\end{equation}
In this case the deceleration parameter is
\begin{eqnarray}
q&=&\frac{1}{2}(1-3c^2).
\end{eqnarray}
To obtain an accelerated expansion universe, i.e. $q<0$, and to
protect the positivity of dark matter energy density, one obtains a
range of values the constant $c$ can take: $1/3<c^2<1$. The
effective equation of state of vacuum energy density is
\begin{eqnarray}
w^{eff}_{\Lambda}&=&-1-\frac{1}{3}\frac{d \ln \rho_{\Lambda}}{d\ln
a}\nonumber\\
&=&-c^2.
\end{eqnarray}
Relation (\ref{rdef}) indicates that the ratio $r \equiv
\rho_m/\rho_{\Lambda} = (1-c^2)/c^2$ is a constant. Although it is
desirable to get a nearly constant $r$ parameter, solving the
coincidence problem, getting a constant value, produces another
problem: there is no transition from decelerated to an accelerated
expansion regime.

\subsection{Variable c}

As we notice at the end of the previous section it is clear that
when $c$ is constant, no transition from decelerated expansion to
accelerated expansion can be realized, a transition observationally
tested. A possible way to overcome this difficulty was proposed in
\cite{pav-zim05}. The idea is to consider $c$ as a time dependent
parameter, enabling also that $r = (1-c^2)/c^2$ to be. Since $r$ is
expected to decrease with time, this implies that $c^2$ increases,
keeping the relation $c^2(1+r)=1$.

In this case
\begin{equation}\label{rholt}
\rho_{\Lambda}=3c^2(t)M^2_{P}H^2.
\end{equation}
and also, using the Friedman equation, we can write the relation
\begin{equation}\label{romt}
\rho_{m}=3M^{2}_{P}(1-c^{2}(t))H^{2}.
\end{equation}
Notice that, although the redshift dependence of $\rho_\Lambda$ is
not proportional to $H^2$, no further constraint on $c^2(z)$ is
derived from the Friedman equation, because $\rho_m$ in (\ref{romt})
is defined to satisfy it automatically. Again, it is to protect the
positivity of the energy density of cold dark matter that
$c(t)^{2}<1$ is required. From the conservation equation of cold
dark matter Eq. (\ref{rom}) and the Friedmann equation, one can
write in a flat universe
\begin{equation}
(1+z)\frac{d\ln H (z)}{dz} =
\frac{3}{2}(1+w_{\Lambda}c^{2}(z))\label{eq:hz}.
\end{equation}
To solve Eq.(\ref{eq:hz}), one has to assume a concrete form of the
function $c(z)$. After a simple calculation, one can also write the
deceleration parameter
\begin{equation}
q=\frac{1}{2}(1-3c^2(z)).
\end{equation}
One finds that once $0<c^{2}(z)<1$ is time dependent, possible
transition from deceleration expansion to accelerated expansion can
be realized. The analogous expression for the effective equation of
state in this case, can be derived using Eqs. (\ref{friedman}) and
(\ref{rholt}), obtaining
\begin{eqnarray}\label{weff2}
w^{eff}_{\Lambda}&=&w_{\Lambda}c^2-\frac{1}{3}\frac{d\ln c^{2}}{d\ln
a}.
\end{eqnarray}
Combining this equation with (\ref{weff}) we obtain
\begin{equation}\label{zimpav13}
w_{\Lambda} = -\left(\frac{1+r}{r} \right)\left[
\frac{Q}{3H\rho_{\Lambda}} + \frac{(c^2)^{\cdot}}{3H c^2}\right],
\end{equation}
a relationship first written in \cite{pav-zim05}, that shows how the
equation of state parameter depends explicitly on the interaction
and the $c^2(t)$ function.

If we write $Q=\Gamma \rho_{\Lambda}$, with $\Gamma$ being some
particle production rate, we can write a solution for (\ref{rom})
\begin{equation}\label{solrom}
\rho_m = \rho_m^0\left(\frac{a_0}{a} \right)^3\exp{\left[\int
 \frac{\Gamma}{r}dt\right]},
\end{equation}
where the subscript indicates today. An explicit model for a flat
universe and interaction was proposed in \cite{zim-pav07}, where it
was suggested that
\begin{equation}\label{ansatz}
\frac{\Gamma}{rH} = 3\beta \left(\frac{a}{a_0} \right)^{\alpha},
\end{equation}
in such a way that (\ref{solrom}) can be solved exactly. In this
case
\begin{equation}\label{roma}
\frac{\rho_{m}}{\rho_{m}^0} = \left(\frac{a_0}{a} \right)^3
\exp{\left[
\frac{3\beta}{\alpha}\left(\left(\frac{a}{a_0}\right)^\alpha - 1
\right) \right]}.
\end{equation}
If we assumed that $c^2$ is constant, then $\rho_\Lambda$ evolve in
the same way as $\rho_m$, so we can explicitly write down the Hubble
parameter from the Friedman equation. The deceleration parameter can
also be computed being
\begin{equation}\label{qzp}
q = \frac{1}{2} - \frac{3}{2} \beta \left( \frac{a}{a_0}
\right)^{\alpha}.
\end{equation}
An identical expression for $q$ was derived in \cite{xu09}, where
instead a explicit form of $c^2(z)$ was given
\begin{equation}\label{c2z}
c^2(z) = \frac{\beta}{(1+z)^{\alpha}}.
\end{equation}
The equivalence between the models can be easily explained using the
equation (\ref{zimpav13}). The analysis performed in \cite{xu09}
consider a flat universe $k=0$ with interaction and is explicitly
used that $w_{\Lambda}=-1$. This last assumption implies a relation
between the interaction $Q$ and $c^2$. Actually, up to a constant,
the relation is $c^2=Q/(3rH\rho_\Lambda)$. The analysis performed
both in \cite{zim-pav07} and \cite{xu09} correspond to this model,
considering interaction between the dark constituents in a
\textit{flat} universe, keeping the value $c^2(1+r)=1$.

\section{Non flat model}

In this section we proceed to generalize the previous work to non
flat universes. Although it is often said that the inflationary
scenario predicts a flat universe \cite{inflation}, the situation
has not been settled down yet, partly because the observations are
not conclusive and because in theory $\Omega_k = 0$
($\Omega_k=-k/a^2H^2$ in this paper) is not the only solution
\cite{lin-mez}. Moreover, in the context of parameter estimations,
considering the curvature as a new free parameter, leads to an
increment in the uncertainties in all the others cosmological
parameters. In particular, in the context of models with dynamical
dark energy, a number of works have discussed the degeneracies
between $w(z)$ and $\Omega_k$ \cite{linder05}, \cite{polran},
\cite{clcoba}. Therefore, the precision in observational cosmology
requires the inclusion of the curvature, even if its contribution is
small, in any theoretical study of dark energy. In this section
$c^2$ is a function of time and $\Omega_k \neq 0$.

Let us start with Eq.(\ref{hdot}). Replacing the definitions of both
$\rho_m$ and $\rho_{\Lambda}$ we found
\begin{eqnarray}\label{hdot2}
\dot{H}=-{3 \over 2}H^2(1 + w_{\Lambda}c^2) - {k \over 2a^2}.
\end{eqnarray}
Using that $a=(1+z)^{-1}$ we can rewrite this equation using the
redshift as the independent variable
\begin{eqnarray}\label{hz2}
(1+z)\frac{dH^{2}}{dz} = 3H^{2}(1+w_{\Lambda}c^{2})+ k (1+z)^2,
\end{eqnarray}
which is the generalization of (\ref{eq:hz}). The generalization of
(\ref{zimpav13}) follows in the same way as
\begin{eqnarray}
w_{\Lambda} = -\left(\frac{1+r}{r+\Omega_k} \right)\left[
\frac{Q}{3H\rho_{\Lambda}} + \frac{\Omega_k}{3}+
\frac{(c^2)^{\cdot}}{3H c^2}\right],
\end{eqnarray}
that contains the obvious limit $\Omega_k=0$ result
(\ref{zimpav13}), and also coincides with the result given in
\cite{zim-pav07} (see equation (26) in that paper). To show the
equivalence we use the identities $\Omega_m + \Omega_k +
\Omega_{\Lambda}=1$ and $r=\Omega_m/\Omega_{\Lambda}$. The analog of
equation (\ref{weff2}) in this case shows how the presence of
curvature changes the dynamics. The effective equation of state
parameter when $w_\Lambda=-1$ takes the form
\begin{eqnarray}\label{weff3}
w^{eff}_{\Lambda}&=&-c^2-\frac{\Omega_k}{3} -\frac{1}{3}\frac{d\ln
c^{2}}{d\ln a}.
\end{eqnarray}
The presence of curvature can ameliorate the negativity of
$w^{eff}_{\Lambda}$ or, if it is positive, it can be even more
negative, crossing the phantom line. However, the statistical
analysis performed on this model, that we describe below, shows that
because the curvature parameter turns out to be very small (see
results in Table I), the actual value of $\Omega_k$ has a little
impact on the dynamics of $w^{eff}_{\Lambda}$ (see Figure
\ref{fig2}).

An explicit solution of (\ref{hz2}) is possible when we fix the bare
equation of state parameter $w_{\Lambda}=-1$
\begin{eqnarray}\label{hz3}
H^{2}(z)=H_0^{2}(1+z)^{3} e^{\frac{3c^2}{\alpha}} \times \\
\nonumber \left[ e^{-\frac{3\beta}{\alpha}} + \frac{\Omega_k}{\beta}
\left( \frac{3\beta}{ \alpha} \right)^{-\frac{1}{\alpha}} \left\{
\Gamma({1 \over \alpha}, {3c^2 \over \alpha}) -\Gamma({1 \over
\alpha}, {3\beta \over \alpha}) \right\} \right],
\end{eqnarray}
where $\Gamma(a,z)$ is the incomplete gamma function or plica
function defined as
\begin{eqnarray}
\Gamma(a,z) = \int^{\infty}_{z} t^{a-1}e^{-t}dt.
\end{eqnarray}
Expression (\ref{hz3}) for $\Omega_k=0$ reduces to the expression
first written in \cite{zim-pav07} and contrasted with observations
in \cite{xu09}.

Having the Hubble parameter in terms of the redshift, we can compute
the luminosity distance and confront the model with observations of
supernovae (SN) \cite{ref:SNchi2}. In this work, we use the
Supernova Cosmology Project Union sample \cite{SCP}, having $307$ SN
distributed over the range $0.015 < z < 1.551$. We fit the
(theoretical) distance modulus $\mu(z)_{th}$ defined by
\begin{equation}\label{muz}
\mu(z)_{th} = 5\log_{10}\left[ \frac{H_0 d_L(z)}{c}\right]+\mu_0,
\end{equation}
to the observational ones $\mu(z)_{obs}$. Here $H_0=100 h$km
s$^{-1}$ Mpc$^{-1}$ is the Hubble constant and the luminosity
distance is defined by $ d_L(z)=(1+z)r(z)$ where
\begin{equation}
r(z) = \frac{c}{H_0 \sqrt{ \left| \Omega_k \right|}} \text{Sinn}
\sqrt{ \left| \Omega_k \right|} \int_0^z \frac{dz'}{H(z')},
\end{equation}
and $ \mu_0  =  42.38 - 5\log_{10}h $. Sinn$(x)=\sin x, x, \sinh x$
for $\Omega_k<0$, $\Omega_k=0$, and $\Omega_k>0$ respectively.
Considering $\alpha$, $\beta$ and $\Omega_k$ as free parameters, we
get a $\chi_{min}^2= 310.866$ with the best fit values displayed in
Table I.
\begin{table}
\caption{\label{tab:table1} The best fit values for the free
parameters in the case of a non flat holographic dark universe.
$\alpha$ and $\beta$ characterize the function $c^2(z)$ defined in
(\ref{c2z}), and $\Omega_k$ is the curvature parameter. We specify
the data set used in each .}
\begin{ruledtabular}
\begin{tabular}{ccccc}
Data Set & $\chi^2_{min}$ & $\alpha \pm 0.02$ & $\beta \pm 0.08$ & $\Omega_k \pm 0.009$ \\
\hline
SN & 310.866 & 1.089 & 1.513 & -0.5873 \\
SN+BAO & 310.406 & 0.732 & 3.161 & 0.3876  \\
SN+BAO+CMB & 312.848 & 0.728 & 1.219 & -0.0449 \\
\end{tabular}
\end{ruledtabular}
\end{table}

The second major input for parameter determination comes from the
baryon acoustic oscillations (BAO) detected by Eisenstein et al.
\cite{ref:Eisenstein05}. In our work, we add the following term to
the $\chi^2$ of the model:
\begin{equation}
\chi^2_{BAO} = \left[ \frac{(A-A_{BAO})}{\sigma_A}\right]^{2},
\label{chiBAO}
\end{equation}
where $A$ is a distance parameter defined by
\begin{equation}
A = \frac{\sqrt{\Omega_m H_0^2}}{cz_{BAO}} \left[r^2(z_{BAO})
\frac{cz_{BAO}}{H(z_{BAO})})\right]^{1/3},
\end{equation}
and $A_{BAO}=0.469$, $\sigma_A = 0.017$, and $z_{BAO}=0.35$.
Considering both supernovae and BAO data, we obtain $\chi_{min}^2=
310.406$ with the best fit values displayed in the second row of
Table I.
\begin{figure}[h]
\centering \leavevmode \epsfysize=8cm \epsfbox{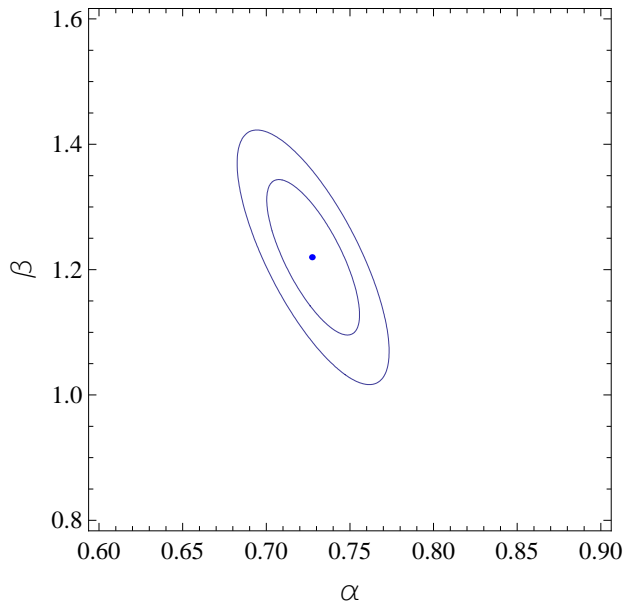} \\
\caption{\label{fig} These are the confidence contours in the plane
$\alpha - \beta$ with $1 \sigma$ and $2 \sigma$ regions. The central
dot shows the best fit parameter for the joint analysis using SN +
BAO + CMB indicated in the third line of table I. This case
correspond to a curvature parameter of $\Omega_k =-0.0449 \pm
0.009$. }
\end{figure}

The CMB shift parameter $R$ is given by \cite{ref:Bond1997}
\begin{equation}
R(z_{\ast})=\sqrt{\Omega_m H^2_0}r(z_{\ast})
\end{equation}
Here the redshift $z_{\ast}$ (the decoupling epoch of photons) is
obtained by using the fitting function \cite{Hu:1995uz}
\begin{equation}
z_{\ast}=1048\left[1+0.00124(\Omega_bh^2)^{-0.738}\right]\left[1+g_1(\Omega_m
h^2)^{g_2}\right],
\end{equation}
where the functions $g_1$ and $g_2$ are given as
\begin{eqnarray}
g_1&=&0.0783(\Omega_bh^2)^{-0.238}\left(1+ 39.5(\Omega_bh^2)^{0.763}\right)^{-1}, \\
g_2&=&0.560\left(1+ 21.1(\Omega_bh^2)^{1.81}\right)^{-1}.
\end{eqnarray}
The WMAP-7 year CMB data alone yields $R(z_{\ast})=1.726\pm0.018$
\cite{wmap7}. Defining the corresponding $\chi^2_{CMB}$ as
\begin{equation}
\chi^2_{CMB}=\frac{(R(z_{\ast})-1.726)^2}{0.018^2}\label{eq:chi2CMB}.
\end{equation}
one can deduce constraints on $\alpha, \beta$ and $\Omega_k$. A
joint analysis using SN+BAO+CMB leads to the best fit values showed
in Table I. As is well known, we notice the effect of incorporating
the BAO and CMB data to the supernova, leading to a decreasing value
for $\Omega_k$. As in the $\Lambda$CDM model, when we use only the
SNIa data, in particular the Union compilation, the fit suggest a
huge curvature contribution (see figure 11 in \cite{SCP} and the
discussion). Only when we incorporate BAO and CMB constraint, the
curvature parameter tends to smaller values. As is also well known,
the strongest observational indication for a flat universe is the
CMB data. A gaussian approximation to the confidence contour is also
shown in Figure \ref{fig}, where we show the plane $\alpha - \beta$
for $\Omega_k=-0.0449$. Using the best fit parameters we show in
Figure \ref{fig2} a plot of the resulting effective equation of
state (\ref{weff3}).

\begin{figure}[h]
\centering \leavevmode \epsfysize=5cm \epsfbox{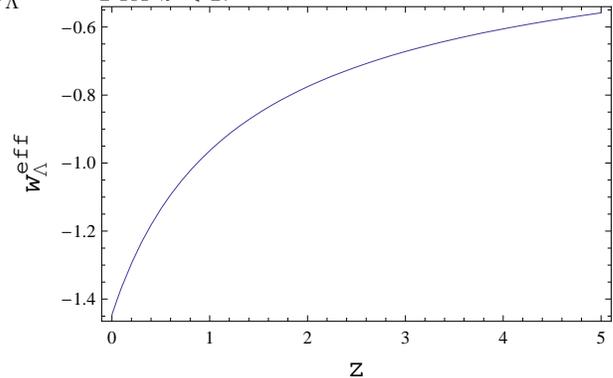} \\
\caption{\label{fig2} Using the best fit parameters found in the
analysis (see Table I), we plot the effective equation of state,
expression (\ref{weff2}), for our model. Notice that even if the
bare equation of state parameter $w_\Lambda =-1$, the combined data
is consistent with an effective equation of state that cross the
cosmological constant line towards phantom like models.}
\end{figure}

By choosing a cosmological constant equation of state $w_\Lambda
=-1$, we get a model that, in agreement with observations, seems to
cross the phantom line, leading to $w^{eff}_\Lambda =-1$ for $z<1$.

\section{Discussion}

We consider a non flat universe version of the holographic dark
energy model studied first in \cite{zim-pav07} and \cite{xu09}, in
which the interaction between dark matter and dark energy enable us
to describe the transition from a decelerated to an accelerated
expansion regime, together with alleviating the cosmic coincidence
problem. Taking into account the spatial curvature, we were able to
find an analytic expression for the Hubble parameter, which was used
to perform a direct confrontation to observational data as SNIa, BAO
and CMB shift parameter. This analysis shows that our model is
compatible with current precision data. The best fit value for the
curvature obtained, although small, it is well within the order
where its presence can modify for example, the number of e-folding
of inflation which changes dramatically the early stages of
inflation \cite{GEllis}.


\begin{thebibliography}{*}

\bibitem{Riess98} A.G. Riess, {\it et al.}, Astron. J. 116
1009(1998) [astro-ph/9805201].

\bibitem{Perlmuter99} S. Perlmutter, {\it et al.}, Astrophys. J. 517
565(1999) [astro-ph/9812133].

\bibitem{bekenstein} J. D. Bekenstein, Phys. Rev. D 7, 2333 (1973);
Phys. Rev. D 23, 287 (1981).

\bibitem{thooft} G. 't Hooft, [gr-qc/9310026], Published in Salamfestschrift:
a collection of talks, eds. A. Ali, J. Ellis and S. Randjbar-Daemi
(Worl Scientific).

\bibitem{susskind} L. Susskind, J. Math. Phys. {\bf 36}, 6377 (1995).

\bibitem{HPple} W. Fischler and L. Susskind, [hep-th/9806039];
D. Bak and S.-J. Rey, Class. Quant. Grav. 17, L83 (2000); R. Tavakol
and G. Ellis, Phys. Lett. B469, 37 (1999); R. Easther and D. Lowe,
Phys. Rev. Lett. 82, 4967 (1999).

\bibitem{cohen} A. Cohen, D. Kaplan and A. Nelson, Phys. Rev. Lett. {\bf 82} (1999) 4971.

\bibitem{thomas} S. Thomas, Phys. Rev. Lett. {\bf 89} (2002) 081301.

\bibitem{hsu} S.D.H. Hsu, Phys. Lett. {\bf B594} 13(2004).

\bibitem{li} M. Li, Phys. Lett. {\bf B603} 1(2004).

\bibitem{pav-zim05} D. Pavon and W. Zimdahl, Phys. Lett. {\bf B628} 206 (2005).

\bibitem{interaction}  B. Wang, C.Y. Lin, E.o Abdalla, Phys. Lett. B 637
357(2006); H. Kim, H.W. Lee, Y.S. Myung, Phys. Lett. B 632
605(2006); B. Hu, Y. Ling, Phys. Rev. D 73 123510(2006); H.M.
Sadjadi, JCAP0702 026(2007);  M.R. Setare, E.C. Vagenas, Int. J.
Mod. Phys. D 18 147(2009); Q. Wu, Y. Gong, A. Wang, J.S. Alcaniz,
[arXiv:0705.1006]; J.F. Zhang, X. Zhang, H.Y. Liu, Phys. Lett. B 659
26(2008); C. Feng, B. Wang, Y. Gong, R.K. Su, [arXiv:0706.4033];
S.F. Wu, P.M. Zhang, G.H. Yang, Class. Quan. Grav. 26 055020(2009).

\bibitem{zim-pav07} W. Zimdahl and D. Pavon, Class. Quant. Grav. {\bf 24}, 5461 (2007); D. Pavon, J. Phys. A{\bf 40}, 6865 (2007).

\bibitem{xu09} L. Xu, JCAP0909, 016 (2009).

\bibitem{GEllis} J-P.Uzan, U. Kirchner, G.F.R. Ellis, Mon. Not. Roy. Astron. Soc. 344, L65
(2003).

\bibitem{inflation} V. Mukhanov, {\it Physical Foundations of
Cosmology}, Cambridge (2005).

\bibitem{lin-mez} A.Linde and A. Mezhlumian, Phys. Rev. D {\bf 52}, 6789
(1995); A. Linde, D. Linde, A. Mezhlumian, Phys. Lett. B{\bf 345},
203,(1995).

\bibitem{linder05} E.V. Linder, Astropart. Phys. {\bf 24}, 391 (2005).

\bibitem{polran} D. Polarski and A. Ranquet, Phys. Lett. {\bf B627}, 1 (2005).

\bibitem{clcoba} C. Clarkson, M. Cortes and B. Bassett, JCAP {\bf 0708}, 011 (2007).

\bibitem{SCP} M. Kowalski et al., Astrophys. J. 686, 749(2008) [arXiv:0804.4142].

\bibitem{ref:SNchi2} S. Nesseris and L. Perivolaropoulos, Phys. Rev. D 72, 123519
(2005) [arXiv:astro-ph/0511040]; L. Perivolaropoulos, Phys. Rev. D
71, 063503 (2005) [arXiv:astro-ph/0412308]; S. Nesseris and L.
Perivolaropoulos, JCAP 0702, 025 (2007) [arXiv:astro-ph/0612653].

\bibitem{ref:Eisenstein05} D. J. Eisenstein, et al, Astrophys. J. 633, 560 (2005) [astro-ph/0501171].

\bibitem{ref:Bond1997} J. R. Bond, G. Efstathiou,
and M. Tegmark, MNRAS 291 L33(1997).

\bibitem{Hu:1995uz} W. Hu, N. Sugiyama, {\it Astrophys. J.} 471 542(1996)
[astro-ph/9510117].

\bibitem{wmap7} E. Komatsu, et.al.,[WMAP Collaboration], arXiv:1001.4538 [astro-ph.CO].



\end{thebibliography}
\end{document}